\documentstyle[12pt,color,epsfig]{article}
\begin{document}

\title{Income Distribution Dependence of Poverty Measure: A Theoretical
Analysis}
\author{${}^{1}$Amit K Chattopadhyay and ${}^{2}$Sushanta K Mallick \\
${}^{1}$Mathematics Institute, University of Warwick \\
Coventry CV4 7AL, UK \\
${}^{2}$ Department of Economics, Loughborough University \\
Loughborough, Leicestershire LE11 3TU, UK}
\maketitle

\begin{abstract}

\noindent
With a new deprivation (or poverty) function, in this paper, we
theoretically study the changes in poverty 
with respect to the `global' mean and variance of the income
distribution using Indian survey data. We show that when the income obeys a 
log-normal distribution,
a rising mean income generally indicates a reduction in poverty 
while an increase in
the variance of the income distribution increases poverty. This altruistic
view for a developing economy, however, is not tenable anymore once the
poverty index is found to follow a pareto distribution. Here although a
rising mean income indicates a reduction in poverty, due to the presence of
an inflexion point in the poverty function, there is a critical value of the
variance below which poverty decreases with increasing variance while beyond
this value, poverty undergoes a steep increase followed by a decrease with
respect to higher variance. Following these results, we make quantitative
predictions to correlate a developing with a developed economy. \newline



\end{abstract}

\newpage

\section{Introduction}

Since the paradigmatic contribution of Sen \cite{Sen1973,Sen1979} and
Atkinson \cite{Atkinson1987}, a remarkable amount of effort has been
undertaken \cite{developing_economics1,developing_economics2} in
theoretically understanding the economics of poverty and inequality. The
studies range from being aptly mathematical in nature to a qualitative
characterisation of such population dialectics. Pradhan and Ravallion \cite%
{Ravallion} have used qualitative assessments of perceived consumption
adequacy based on a household survey. They claim that perceived consumption
needs can be a more promising approach than the subjective income-based
poverty line. This consumption norm can correspond to a saturation level of
consumption, below which the individual could be considered to be in
poverty. Further, in this paper, our approach is rather complementary to a
lemma-based mathematical model in that we use survey based consumption data
to quantify the dependence of a well-known poverty function \cite%
{poverty_function_paper1,poverty_function_paper2} on the mean and variance
of the income distribution. To this end, we use income-expenditure data from
a `developing nation' (India in our case) and utilise the well established
technique of data fitting to define the per capita consumption as a function
of income. Here the implicit assumption is that of a near equilibrium
situation such that the time dependence of both income and consumption
variables can be considered as transients without much effect on the
asymptotic distributions. Deaton \cite{Deaton} has discussed the ambiguity
that arises using survey data versus national accounts data for individual
consumption or income levels. Although the survey consumption data seem to
understate the true consumption levels, we are however using the data as a
backup to our analytical results thereby restricting our claims to being
qualitative in nature. Such comparisons with real data help us have
approximate ideas of the values of the unknown parameters, two in our model,
although the general conclusions are remarkably independent of these
parameter values. \\

\noindent
Assuming that the income distribution can be characterised by a
two-parameter function, such as a log-normal distribution, in the first
section of the paper we study the effects of changes in the mean and
variance of the underlying income distribution on poverty. The results of
this analysis indicate that an increase in mean income and a reduction in
the variance of income distribution can reduce poverty. It also hints
towards a trade-off, in that while an increase in average income reduces
poverty, a simultaneous increase in income variance can escalate poverty.
This result is likely to suggest that reducing income inequality should be
the precondition for lowering poverty. These general results are then
contrasted in the following section using a different model for the income
distribution, the pareto distribution. The objective is basically to probe
whether the results obtained are universal in nature and if not, then which
distribution defines a better measure of poverty. \footnote{%
Sen (1976) introduced the notion of deprivation in the income distribution
literature, and criticised the use of the head-count ratio as a measure of
poverty. Rao (1981) suggested broadening the scope of poverty measurement to
nutritional norms as opposed to monetary measures. If poverty is to be
regarded as negative welfare, it makes sense to relate it to consumption
deprivation resulting from an uneven income distribution rather than to the
income distribution alone as is done by the traditional poverty ratio index
(Kumar et al., 1996).}

\section{Poverty impact of changes in log-normal income distribution}

Poverty equals consumption deprivation on an essential food. The necessity
of defining poverty as a multidimensional concept rather than relying on
income or consumption expenditures per capita has been well documented.
Although it is important to assess deprivation with more than one attribute
(see \cite{Atkinson2003,Bourguignon_Chakravarty2003,
poverty_function_paper1,Mukherjee2001}), we consider the case of most
essential food item that is required for survival, in an attempt to include
deprivation into the poverty index. Such an index would suggest that a
person can be considered poor if the individual's consumption falls within
the deprivation area in the diagram (see lower panel of Fig.1), that is, the
cumulative difference between the saturation consumption level of cereal and
actual cereal consumption by the community as a whole. The upper panel of
Fig.1 shows positive consumption even at zero income level, which makes our
formulation more realistic than the non-linear function used in Kumar et al. 
\cite{Kumar1996}. The non-linear function used in our paper that allows a
saturation level of consumption norm for food-grains is as follows:

\begin{equation}
C(y) = \frac{V \exp(y)}{K+\exp(y)}  \label{consumption_equation}
\end{equation}

\noindent where $C$ is the consumption expenditure on food-grains, $y$ is
income and the parameters $V,\:\:K(>0)$ represent the saturation level of
real food-grain consumption expenditure or the bliss level and the level of
income needed to consume one half of the saturation level respectively.
Consumption deprivation (CD) or poverty (P) can be defined as the shortfall
of actual consumption expenditure relative to saturation level V, or $CD=V-C$%
. Thus the non-linear CD function is derived as:

\begin{equation}
CD=\frac{V K}{K+\exp(y)}  \label{consumption_deprivation_equation}
\end{equation}

\noindent This function, being a convex decreasing function of income
provides a direct measure of poverty based on nutritional norms, while $V$
and $K$ are parameters of a concave Engel curve. Here $C \rightarrow V$
represents the idealistic limit where there is no deprivation or poverty
corresponding to a static equilibrium in the social dialectics
mathematically represented by $y=y^{*}$. In what follows, we would consider
two asymptotic regimes - $y \rightarrow 0$ and $y \rightarrow \infty$ -
physically which correspond to the low and high income groups respectively.
Naturally our focus would be on the $y \rightarrow 0$ limit, that is on the
low income section although the analysis would encompass both limits.

\begin{figure}[tbp]
\centerline{\epsfig{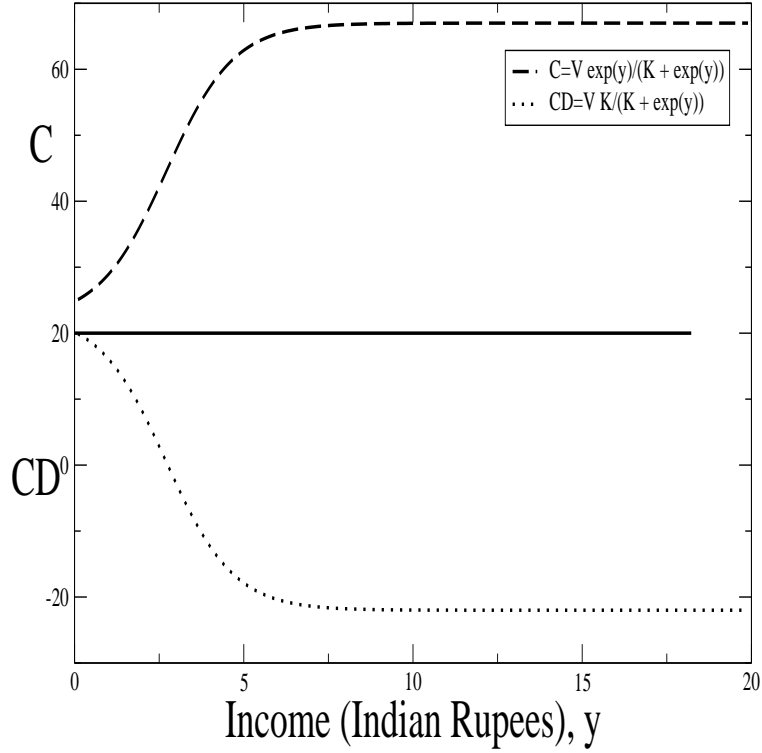}}
\caption{Consumption C(y) and deprivation CD(y) functions against income y}
\label{fig1}
\end{figure}

\noindent If consumption of the most essential food item follows a concave
non-linear functional form and if individual poverty is measured as the
difference between the saturation level of consumption of the essential food
item and its actual level, assumption of a log-normal income distribution
implies a reduction in poverty with the increase of mean income of the
population and an increase in inequality with increasing poverty. This new
measure of poverty is based on the notion of consumption deprivation of a
very essential staple food such as rice or wheat (cereal), derived from a
nonlinear, monotonically increasing concave consumption function varying
with the income, albeit with no specific reference to a subjective poverty
line. The standard log-normal probability density function (pdf) is defined
as

\begin{equation}
{\rm f}(y/\mu,\sigma^2)= \frac{1}{y \sigma \sqrt{2 \pi}}\: \exp[-\frac{{(ln y
- \mu)}^2}{2 {\sigma}^2}]  
\label{lognormal_distribution}
\end{equation}

\noindent where y is normally distributed with mean $\mu$ and variance $%
\sigma^2$ (both positive real numbers). With this log-normal pdf for the
income y, the poverty equation can be rewritten as follows:

\begin{eqnarray}
P &=&\int_{0}^{\infty }\:CD(y)\:{\rm f}(y)dy  \nonumber \\
&=&\int_{0}^{\infty }\:\frac{VK}{K+\exp (y)}\:\frac{1}{y\sigma \sqrt{2\pi }}%
\:\exp [-\frac{{(lny-\mu )}^{2}}{2{\sigma }^{2}}]\:dy
\label{poverty_equation}
\end{eqnarray}

\noindent Partial derivatives of the above equation (\ref{poverty_equation})
with respect to $\mu$ and $\sigma^2$ give

\begin{eqnarray}
\frac{\partial P}{\partial \mu} &=& \int_{0}^{\infty}\:\frac{V K}{K +
\exp(y) }\:\frac{1}{y \sigma \sqrt{2 \pi}}  \nonumber \\
&\times& \exp[-\frac{{(ln y - \mu)}^2}{2 {\sigma}^2}] \frac{ln y - \mu}{
\sigma^2}\:dy  \nonumber \\
\frac{\partial P}{\partial {\sigma^2}} &=& \int_{0}^{\infty}\:\frac{V K}{K +
\exp(y)}\:\frac{1}{2 y {\sigma^3} \sqrt{2 \pi}}  \nonumber \\
&\times& \exp[-\frac{{(ln y - \mu)}^2}{2 {\sigma}^2}]\: [\frac{{(ln y - \mu)}
^2}{\sigma^2}-1]\:dy  \label{derivative_equations}
\end{eqnarray}

\noindent P satisfies the three standard axioms of a poverty index \footnote{%
See Foster {\em et al} \cite{developing_economics1}, Kakwani \cite%
{developing_economics2}, Atkinson \cite{Atkinson1987}, and Foster and
Shorrocks \cite{poverty_function_paper2} for the different axioms of a
poverty index.}, namely the monotonicity, transfer, and transfer sensitivity
axioms that any such index must satisfy.

\subsection{Asymptotic solutions of the poverty function}

\noindent This section deals with asymptotic solutions of the poverty
functions for extremely low ($y\rightarrow 0$) to moderate values of the
income distribution. This is mathematically categorised in the following
manner: \newline

\noindent For moderate incomes, $C(y)=\frac{V \exp(y)}{K+\exp(y)}=C_{{\rm mod%
}}(y)$, say, \newline

\noindent 1. $C_{{\rm mod}}(y \rightarrow 0) = \frac{V}{K+1}$ $\&$ \newline
\noindent 2. $C_{{\rm mod}}(y \rightarrow \infty) = V$

\noindent whereas for very low income groups, $C(y)=\frac{V y}{K+y}=C_{{\rm %
\ low}}(y)$, say,\newline

\noindent 1. $C_{{\rm low}}(y \rightarrow 0) = 0$ $\&$ \newline
\noindent 2. $C_{{\rm low}}(y \rightarrow \infty) = V$ \newline

\noindent The above comparison clearly shows that although both definitions
of the consumption function are generally equivalent in the low income
limit, for the absolutely needy groups, $C_{{\rm mod}}(y)$ predicts a
non-zero ($\frac{V}{K+1}$) lower limit of income which is more realistic
than $C_{{\rm low}}(y\rightarrow 0)=0$. A linear stability analysis of $C_{%
{\rm low}}(y)$ also shows that $y=0$ is an unstable fixed point, which
further strengthens this conviction. Henceforth our attention will mainly be
focused towards the lowest income groups defined by $C_{{\rm low}}(y)$,
although we would flip back and forth between the moderate to the low income
classes for comparisons. Before proceeding any further, though, we first
derive the values of the parameters $V,K$ by fitting the function $C_{{\rm %
mod}}$ with actual survey data obtained from National Sample Survey,
1999-2000, 55th Round, India. We would be using these values of $V,K$ in all
analyses in this paper. Fig.2 portrays the shape of an Engle curve, graphing
real cereal expenditure against the total expenditure - a surrogate for
income.\newline

\begin{figure}[tbp]
\centerline{\epsfig{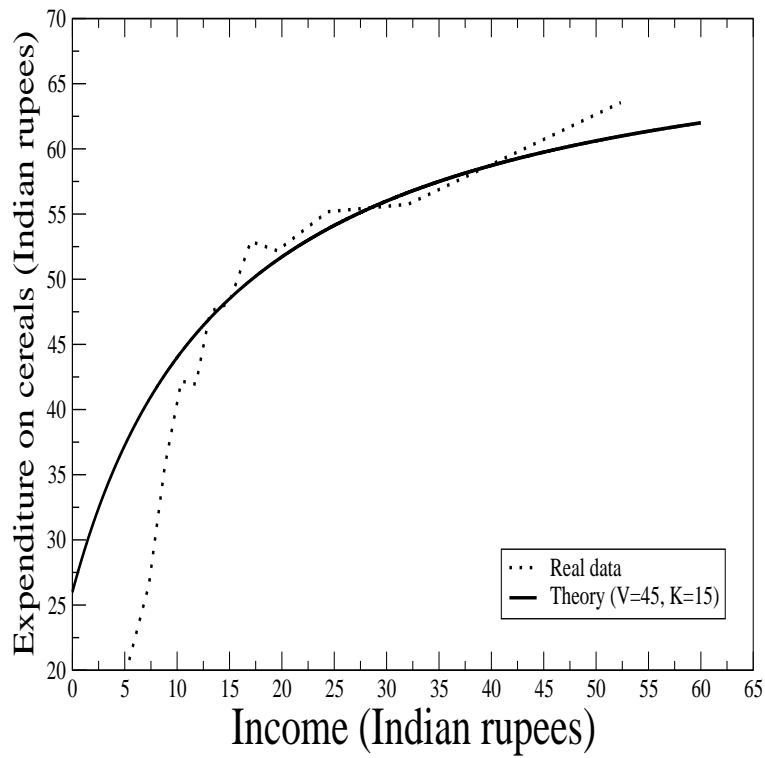}}
\caption{Consumption C(y) plotted against income y: data fitting to evaluate
V and K using data from Indian National Sample survey 1999-2000, 55${}^{{\rm %
th}}$ round.}
\label{fig2}
\end{figure}

\noindent The above exact data fitting conclusively shows that the
parameters $V$ and $K$ have the respective values 45 and 15 in Indian
currency (Rupees). These are roughly equivalent to 1.0 USD and 0.33 USD
respectively. Now using these values, we study the case for typically the
lowest income classes defined by the consumption function $C_{{\rm low}}(y)$%
. In this case, however, we need to focus on both low and high limits of the
variance. Upto first order in $\sigma ^{2}$, we find that

\begin{equation}
{P_{{\rm low}}}_{\sigma \rightarrow 0} = \frac{V}{K} [K - \exp(-{\mu - \frac{%
\sigma^2}{2}})]  \label{lognormal_low_small_sigma}
\end{equation}

\noindent The poverty dependence on the mean for this asymptotic regime can
be understood from figure 3. \\

\begin{figure}[tbp]
\centerline{\epsfig{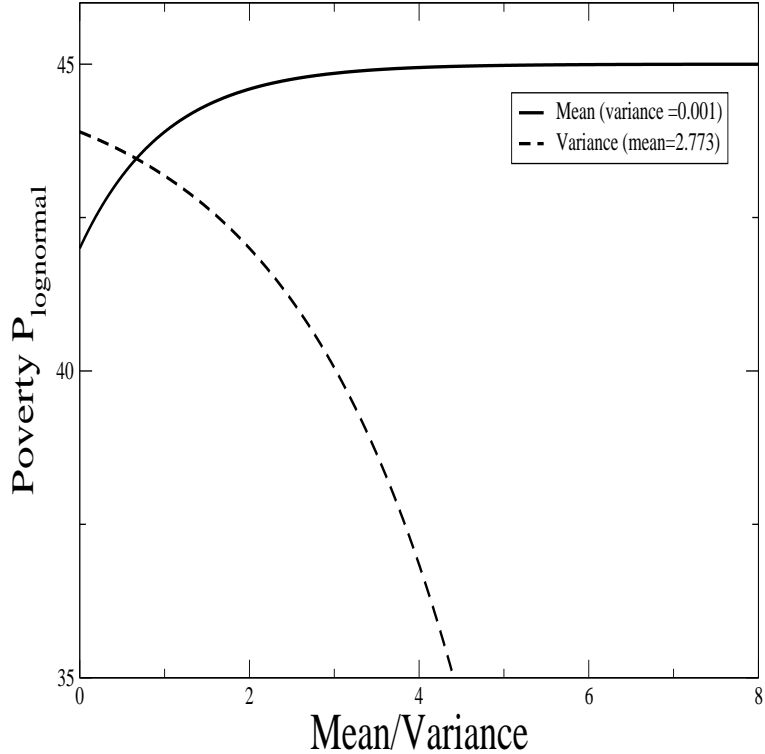}}
\caption{Poverty versus mean $\protect\mu$ (for fixed variance $\protect%
\sigma^2 =0.001$) and versus variance $\protect\sigma^2$ (for a fixed mean $%
\protect\mu=2.773$) for a log-normal distribution for the limiting
case $\sigma^2 \rightarrow 0$ defined by equation (\protect\ref%
{lognormal_low_small_sigma}).}
\label{fig3}
\end{figure}

\noindent Fig. 3 tells us that poverty is a monotonically decreasing
function of variance for a fixed mean (taken to be 2.773 for a direct
comparison with Fig. 4 later). On the other hand, for a fixed variance
(0.001), poverty increases with mean and then saturates after a critical
value. This result is very remarkable but needs to be taken with a pinch of
salt, especially since this is true only in the asymptotic ($\sigma
\rightarrow 0$) regime. We will revisit this problem in the following
section where we discuss the situation when both the mean and the variance
of the income distribution are simultaneously varying.

\subsection{Overall impact of simultaneous changes in mean and variance}

Here we show what effect any change, either increase or decrease, in the
income distribution has on the overall poverty function when the
distribution is log-normal and when both mean and variance are varying.
Since our focus is on the low income group, we will be using $C_{{\rm low}%
}(y)$ as our definition for the consumption function. The attention here
would be to decipher the joint variation of the poverty function $P(\mu
,\sigma ^{2})$ with respect to $\mu $ and $\sigma ^{2}$. Once again using a $%
1/y$ expansion \footnote{%
\noindent This might sound confusing since we are discussing small income
but in effect, all that we are doing is to use a well known 1/y expansion
prevalent in statistical mechanics. It is generally valid for a considerable
range involving large to moderate values of the variable y. We have checked
this result using $C_{{\rm mod}}$ and the qualitative results remain
altogether unaltered.} upto the first order, we find that the joint poverty
function reads as

\begin{eqnarray}
dP(\mu,\sigma^2) &=& \frac{\partial P}{\partial \mu} d{\mu} + \frac{\partial
P }{\partial {\sigma^2}} d{\sigma^2}  \nonumber \\
&=& \frac{V K}{\sigma^2} \exp[-(\mu - \frac{\sigma^2}{2})]\:d{\mu}  \nonumber
\\
&+& \frac{V K}{2} \exp[-(\mu - \frac{\sigma^2}{2})A]\:d{\sigma^2}
\label{lognormal_equation}
\end{eqnarray}

\noindent This equation suggests that poverty is a decreasing function of
changes in $\mu $ and an increasing function of changes in $\sigma $$^{2}$.
For a fixed variance, d$\sigma $$^{2}$=0, and hence the first component of
[7], reflecting change in $\mu $, will provide convergence; and with a fixed
mean, d$\mu $=0, the second component, exhibiting change in $\sigma $$^{2}$,
will give convergence of the equation. When both the mean and variance of
the income distribution change as a result of changes in macroeconomic
policies, their effect on poverty can be evaluated via equation (\ref%
{lognormal_equation}). The notable point here is the fundamental qualitative
difference with the prediction from equation (\ref{lognormal_low_small_sigma}%
). As opposed to the earlier asymptotic result where increase of the mean
income was expected to generate a positive augmentation in poverty (for
fixed variance) followed by a saturation at a particular value $\mu _{c}$,
equation (\ref{lognormal_equation}) with a fixed $\sigma $ clearly suggests
that poverty decreases with increase of the mean income. This apparent
dichotomy can be understood once we analyse the physical meaning hidden in
equation (\ref{lognormal_low_small_sigma}). It says that in a relatively
large group of low earning population, a very small variance between the
earners contributes to an increase in poverty for very low to moderate
values of the mean income. However, once the mean income reaches a critical
value, this spurious effect saturates off. This can be contrasted with the
prediction from the last equation which holds true for moderate to large
values of $\sigma $. We would like to specifically point out here that both
predictions from equations (\ref{lognormal_low_small_sigma}, \ref%
{lognormal_equation}) are true but in their respective regimes defined by
small to large values of $\sigma $.

\section{Poverty impact of changes in pareto income distribution}

\noindent In this section, our objective is to study the mean and variance
dependence of the poverty function, replacing the log-normal probability
distribution, previously assumed, with a pareto distribution and contrast
the findings later. Once again we would conform to the same consumption and
deprivation functions (\ref{consumption_equation},\ref%
{consumption_deprivation_equation}) and try to understand the qualitative
changes in the poverty function of a growing economy with respect to changes
in the mean and variance of the overall income distribution.

\noindent The standard pareto probability density function $f_{{\rm pareto}
} $ defined over the interval $y \geq b$ is given by

\begin{equation}
{\rm f}_{{\rm pareto}}(y) = \frac{a b^a}{y^{a+1}}
\label{pareto_distribution}
\end{equation}

\noindent where the mean $\mu$ and the variance $\sigma^2$ can be easily
shown to be as follows

\begin{eqnarray}
\mu &=& \frac{ab}{a-1}  \nonumber \\
\sigma^2 &=& \frac{a b^2}{{(a-1)}^2 (a-2)}  \label{pareto_mean_variance}
\end{eqnarray}

\noindent With this pareto probability density function, the poverty
function $P_{{\rm pareto}}$ reads as follows

\begin{eqnarray}
P_{{\rm pareto}}(a,b) &=& \int_{0}^{\infty}\:CD(y) {\rm f}_{{\rm pareto}}(y)
dy  \nonumber \\
&=& V K a b^a\:\int_{b}^{\infty}\:\frac{dy}{(K+y) y^{a+1}}  \nonumber \\
&=& \frac{V}{K}\:[1 - a b^a \int_{b}^{\infty}\:dy \frac{1}{(K+y) y^a}]
\end{eqnarray}

\noindent Defining the identity $I(a)=\int_{b}^{\infty }\:\frac{dy} {%
(K+y)y^{a+1}}$, and taking recourse to a bit of algebra one can deduce a
recursive relation

\begin{eqnarray}
I(a) &=& \frac{1}{K a b^a}[1 - \frac{a b}{K(a-1)} + \frac{a b^2}{K^2 (a-2)}]
\nonumber \\
&-& \frac{1}{K^3} I(a-3,b)  \label{hypergeometric_series}
\end{eqnarray}

\noindent This equation (\ref{hypergeometric_series}) can be correlated with
a hypergeometric ${}_{2}F_1$ series \footnote{%
A hypergeometric series is an algebraic power series in which the ratio of
successive coefficients $r_n/r_{n-1}$ is a rational function of $n$. The
hypergeometric series that we are using here is due to Gauss and has the
mathematical definition ${}_{2}F_{1}(a,b;c;,z) = \frac{\Gamma(c)}{\Gamma(b)
\Gamma(c-b)} \:\:\int_{0}^{1} dt \:t^{b-1}\:{(1-t)}^{c-b-1}\:{(1-tz)}^{-a}$.
In our case, $I(a)=\frac{b^{-1-a}}{1+a}\:{}_{2}F_{1}(1,1+a;2+a;-K/b)$ for $%
b>0,b+K \geq 0, {\rm Re}[a]>-1, {\rm Im}[K] \neq 0$.} and for specified
values of the parameters can be solved numerically. For our purpose though,
we consider the limit $a \rightarrow \infty$ to have a first hand impression
of the situation

\begin{equation}
P(a \rightarrow \infty,b) = \frac{V}{K}\:[1 - \frac{1}{K} \frac{1}{1+\frac{b 
}{K}}]
\end{equation}

\noindent We would now directly evaluate the poverty function in a more
physical limit. Without any loss of generality we choose the limit $%
K\rightarrow 0$ which is akin to the 1/y expansion we did in deriving the
poverty function for the log-normal distribution. We would shortly see that
in this case, this basic expansion allows us to have an `exact' derivation
of the poverty function as opposed to its log-normal counterpart. Upto the
first order in 1/y and utilising equation (\ref{pareto_mean_variance}), we
find

\begin{equation}
P(a,b) = VK [\frac{1}{\mu} \frac{a^2}{a^2-1} - \frac{1}{\mu^2} \frac{a^3}{{\
(a-1)}^2 (a+2)}]  \label{pareto_poverty_function}
\end{equation}

\noindent where $a=2+\sqrt{1+\frac{\mu ^{2}}{\sigma ^{2}}}$ and $b=\frac{a-1 
}{a}\mu $. A numerical solution of the above equation (\ref%
{pareto_poverty_function}) \footnote{%
To evaluate the inflexion points, we used the software mathematica and later
checked the result using another software called maple. The results were
once again cross-checked using a self-generated fortran code. All numerical
results that we cite in this article have been cross-checked using three
different and independent numerical techniques.} shows that it has a pair of
inflexion points \footnote{%
\noindent The inflexion point is defined through the numerical solution of
the coupled equations $\frac{\partial^2 P}{\partial \mu^2}=0$ and $\frac{%
\partial^2 P}{\partial s^2}=0$, where $s=\sigma^2$. Out of the two pairs of
solution, only one turns out to be physical. The other solution gives a
negative value of $s$. We work with the physical solution only.}, out of
which the physical pair is at $\mu =3.05139\:\:\&\:\:\sigma ^{2}=0.0692138$.
Solving around this inflexion point, we now come across one of the most
remarkable results of this article, the fact that poverty initially
decreases with increasing variance until it reaches a critical value $\sigma
^{2}={\sigma _{c}}^{2}$ beyond which the poverty starts increasing with
variance followed by a dip once again. \par

\begin{figure}[tbp]
\centerline{\epsfig{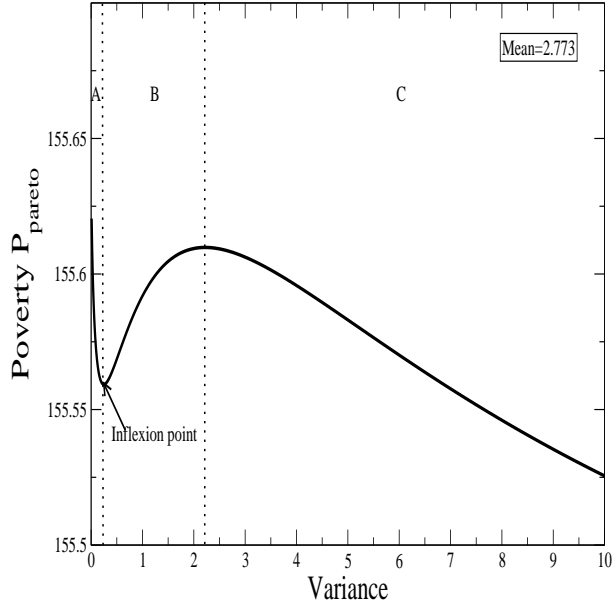}}
\caption{Poverty versus mean $\protect\mu $ for fixed variance in a pareto
distribution close to the inflexion point($\protect\mu =2.773$). Zone {\bf A}
represents an under-developed economy; zone {\bf B} defines the poverty-
variance relation for a developing nation; zone {\bf C} represents 
an economically developed nation.}
\label{fig4}
\end{figure}

\noindent
Fig. 4 has been drawn using $\sigma ^{2}=2.773$, a value reasonably close to
the inflexion point. The plot shows that poverty decreases until it reaches
the point $\mu _{c}\sim 0.25$ after which it starts increasing approximately
until $\sigma ^{2}=2$ and then it starts decreasing again. This result is in
marked contrast with the log-normal case where the poverty rather
uninterestingly decreases with increasing mean for a fixed variance, and
increases with for a fixed mean. It is now not difficult to pinpoint the
detailed meaning of this result. Referring to Fig. 4, zone A defines a
rather `underdeveloped' economy, zone B stands for a `developing' economy,
our case in study, while the final zone C clearly indicates what one would
expect in the case of an economically `developed' nation. We can probably
claim without much ambiguities that a pareto distribution has the power to
encapsulate all three modes of economies and is the ideal candidate for all
future studies involving poverty measure. Further, zones B and C appear to
suggest an inverted-U hypothesis similar to Kuznets \cite{Kuznets} that
poverty increases in the early stages of development and subsequently it
declines with higher level of economic progress even though such development
is associated with higher inequality.

\section{Conclusion}

\noindent This paper made use of a poverty function, which is different from
the conventional poverty indices in the following manner: (1) the CD index
does not depend on an arbitrarily chosen poverty line, (2) it depends on the
observed and measurable consumption behaviour of people, (3) the index
satisfies the standard axioms of a poverty index. Having used such a
consumption deprivation function as a measure of poverty, this paper has
shown analytically that for a log-normal income distribution, an increase in
mean income, ceteris paribus, will decrease poverty while an increase in the
variance of the income distribution, ceteris paribus, will increase poverty
although somewhat contradictory information was obtained for the limiting
case of earners with extremely low variance in their income distribution. In
this case, poverty was found to decrease with increasing variance for a
fixed mean, while when plotted against the mean (Fig. 3), it was found to
initially increase and then saturate after a critical value of the mean
which we could determine theoretically. \\

\noindent
These observations were later contrasted with observations made from a
pareto distribution. Here we found that for very low earning groups in a
developing economy, poverty initially decreases with increasing variance but
beyond a critical value of the variance, it starts increasing later to
decrease again. In the process, this defines all three economies
characterised by individual parametric regimes. The conclusion that we
derive from these joint analyses is that the variance dependence of poverty
is not unequivocally simplistic, in that one distribution (log-normal)
predicts an increase in poverty with increasing variance (although the
limiting $\sigma ^{2}\rightarrow 0$ case was somewhat qualitatively
identical to zone B for the pareto distribution) while the other (pareto)
shows the existence of an inflexion point in the poverty function. This
means that the poverty-variance graph in a pareto distribution has a
critical point, on one side (zone A) of which poverty decreases with
increasing variance, while on the other side it is just the reverse. \\

\noindent
Our contribution here is to prove that a pareto distribution offers the more
realistic measure of poverty in a developing economy. This is because it
condones the very realistic fact that for very low income groups a slight
increase in the variance only serves to decrease poverty whereas for high
earning groups, greater the variation in earning greater is the probability
of an escalation in poverty up to another critical point, beyond which
poverty declines with any further increase in variance of wealth
distribution in a society. This phase seems to reflect the case of a very
developed economy, one which we identify as the supra-economic behaviour. In
macroeconomic sense, this phase suggests that close to an equilibrium
dynamics, higher inequality could contribute to higher savings and thereby
higher growth and reduced poverty. In a follow-up work \cite{future_work},
shortly to be communicated, we have shown that in the non-stationary case,
where both income and consumption are functions of time, the consumption
deprivation dynamics can be mapped to the paradigmatic Burgers' equation, 
\footnote{%
Burger's equation is a 1+1 dimensional equation which generally represents
the time change in the velocity of a fluid flowing under constant pressure}
thereby bestowing us with the ability to make quantitative predictions on
the poverty of a developing economy as a function of income and time.

\end{document}